%% file: main.tex
\documentclass[pdftex,twocolumn,epjc3]{svjour3}          

\RequirePackage[T1]{fontenc}

\smartqed  

\RequirePackage{graphicx}
\RequirePackage{mathptmx}      
\RequirePackage{flushend}
\RequirePackage[numbers,sort&compress]{natbib}
\RequirePackage[colorlinks,citecolor=blue,urlcolor=blue,linkcolor=blue]{hyperref}

\usepackage{multirow}
\usepackage{amsmath}
\usepackage{amssymb}
\usepackage{dblfloatfix} 
\usepackage{hyperref}
\usepackage{microtype} 
\usepackage{graphicx}
\usepackage{dcolumn}
\usepackage{bm}
\usepackage[mathlines]{lineno}
\usepackage{xcolor}
\usepackage{booktabs}
\usepackage{orcidlink}

\usepackage{natbib}
\journalname{Eur. Phys. J. C}

\newcommand{\iso}[2]{$\rm{^{#2}#1}$}

\newcommand{\revised}[1]{\textcolor{black}{#1}}

\DeclareUnicodeCharacter{03B1}{\ensuremath{\alpha}}
\DeclareUnicodeCharacter{03B2}{\ensuremath{\beta}}
\DeclareUnicodeCharacter{03B3}{\ensuremath{\gamma}}

\begin{document}

\title{Nonproportionality of NaI(Tl) Scintillation Detector for Dark Matter Search Experiments}

\input{source/COSINE-100_authors_September2023_v2_epjc}

\date{Received: \today / Accepted: N/A}

\maketitle

\input{source/abstract}
\input{source/introduction}

\input{source/cosine}

\input{source/internal}

\input{source/spectroscopy}

\input{source/result}

\input{source/conclusion}

\begin{acknowledgements}
\input{source/acknowledgments_September2023_v3}
\end{acknowledgements}

\bibliography{source/bibliography}

\end{document}

%% file: source/COSINE-100_authors_September2023_v2_epjc.tex
\author{
    S.~M.~Lee\thanksref{corrauthor0,addr:snu}\orcidlink{0000-0002-7326-3317}\and
    G.~Adhikari\thanksref{addr:yale}\and
    N.~Carlin\thanksref{addr:saopaulo}\and
    J.~Y.~Cho\thanksref{addr:cup}\and
    J.~J.~Choi\thanksref{addr:snu,addr:cup}\and
    S.~Choi\thanksref{addr:snu}\and
    A.~C.~Ezeribe\thanksref{addr:sheffield}\and
    L.~E.~Fran{\c c}a\thanksref{addr:saopaulo}\and
    C.~Ha\thanksref{addr:cau}\and
    I.~S.~Hahn\thanksref{addr:cens,addr:ewu,addr:ibs}\and
    S.~J.~Hollick\thanksref{addr:yale}\and
    E.~J.~Jeon\thanksref{addr:cup,addr:ibs}\and
    H.~W.~Joo\thanksref{addr:snu}\and
    W.~G.~Kang\thanksref{addr:cup}\and
    M.~Kauer\thanksref{addr:wismad}\and
    B.~H.~Kim\thanksref{addr:cup}\and
    H.~J.~Kim\thanksref{addr:knu}\and
    J.~Kim\thanksref{addr:cau}\and
    K.~W.~Kim\thanksref{addr:cup}\and
    S.~H.~Kim\thanksref{addr:cup}\and
    S.~K.~Kim\thanksref{addr:snu}\and
    S.~W.~Kim\thanksref{addr:cup}\and
    W.~K.~Kim\thanksref{addr:ibs,addr:cup}\and
    Y.~D.~Kim\thanksref{addr:cup,addr:sejong,addr:ibs}\and
    Y.~H.~Kim\thanksref{addr:cup,addr:kriss,addr:ibs}\and
    Y.~J.~Ko\thanksref{corrauthor1,addr:cup}\orcidlink{0000-0002-5055-8745}\and
    D.~H.~Lee\thanksref{addr:knu}\and
    E.~K.~Lee\thanksref{addr:cup}\and
    H.~Lee\thanksref{addr:ibs,addr:cup}\and
    H.~S.~Lee\thanksref{addr:cup,addr:ibs}\and
    H.~Y.~Lee\thanksref{addr:cens}\and
    I.~S.~Lee\thanksref{addr:cup}\and
    J.~Lee\thanksref{addr:cup}\and
    J.~Y.~Lee\thanksref{addr:knu}\and
    M.~H.~Lee\thanksref{addr:cup,addr:ibs}\and
    S.~H.~Lee\thanksref{addr:ibs,addr:cup}\and
    Y.~J.~Lee\thanksref{addr:cau}\and
    D.~S.~Leonard\thanksref{addr:cup}\and
    N.~T.~Luan\thanksref{addr:knu}\and
    B.~B.~Manzato\thanksref{addr:saopaulo}\and
    R.~H.~Maruyama\thanksref{addr:yale}\and
    R.~J.~Neal\thanksref{addr:sheffield}\and
    J.~A.~Nikkel\thanksref{addr:yale}\and
    S.~L.~Olsen\thanksref{addr:cup}\and
    B.~J.~Park\thanksref{addr:ibs,addr:cup}\and
    H.~K.~Park\thanksref{addr:koreau}\and
    H.~S.~Park\thanksref{addr:kriss}\and
    J.~C.~Park\thanksref{addr:cnu}\and
    K.~S.~Park\thanksref{addr:cup}\and
    S.~D.~Park\thanksref{addr:knu}\and
    R.~L.~C.~Pitta\thanksref{addr:saopaulo}\and
    H.~Prihtiadi\thanksref{addr:negerimalang}\and
    S.~J.~Ra\thanksref{addr:cup}\and
    C.~Rott\thanksref{addr:skku,addr:utah}\and
    K.~A.~Shin\thanksref{addr:cup}\and
    D.~F.~F.~S. Cavalcante\thanksref{addr:saopaulo}\and
    A.~Scarff\thanksref{addr:sheffield}\and
    M.~K.~Son\thanksref{addr:cnu}\and
    N.~J.~C.~Spooner\thanksref{addr:sheffield}\and
    L.~T.~Truc\thanksref{addr:knu}\and
    L.~Yang\thanksref{addr:saopaulo}\and
    G.~H.~Yu\thanksref{addr:skku,addr:cup}\\
    (COSINE-100 Collaboration)
}

\thankstext{corrauthor0}{e-mail: physmlee@gmail.com}
\thankstext{corrauthor1}{e-mail: yjko@ibs.re.kr}

\institute{
    Department of Physics and Astronomy, Seoul National University, Seoul 08826, Republic of Korea\label{addr:snu}\and
    Department of Physics and Wright Laboratory, Yale University, New Haven, CT 06520, USA\label{addr:yale}\and
    Physics Institute, University of S\~{a}o Paulo, 05508-090, S\~{a}o Paulo, Brazil\label{addr:saopaulo}\and
    Center for Underground Physics, Institute for Basic Science (IBS), Daejeon 34126, Republic of Korea\label{addr:cup}\and
    Department of Physics and Astronomy, University of Sheffield, Sheffield S3 7RH, United Kingdom\label{addr:sheffield}\and
    Department of Physics, Chung-Ang University, Seoul 06973, Republic of Korea\label{addr:cau}\and
    Center for Exotic Nuclear Studies, Institute for Basic Science (IBS), Daejeon 34126, Republic of Korea\label{addr:cens}\and
    Department of Science Education, Ewha Womans University, Seoul 03760, Republic of Korea\label{addr:ewu}\and
    IBS School, University of Science and Technology (UST), Daejeon 34113, Republic of Korea\label{addr:ibs}\and
    Department of Physics and Wisconsin IceCube Particle Astrophysics Center, University of Wisconsin-Madison, Madison, WI 53706, USA\label{addr:wismad}\and
    Department of Physics, Kyungpook National University, Daegu 41566, Republic of Korea\label{addr:knu}\and
    Department of Physics, Sejong University, Seoul 05006, Republic of Korea\label{addr:sejong}\and
    Korea Research Institute of Standards and Science, Daejeon 34113, Republic of Korea\label{addr:kriss}\and
    Department of Accelerator Science, Korea University, Sejong 30019, Republic of Korea\label{addr:koreau}\and
    Department of Physics and IQS, Chungnam National University, Daejeon 34134, Republic of Korea\label{addr:cnu}\and
    Department of Physics, Universitas Negeri Malang, Malang 65145, Indonesia\label{addr:negerimalang}\and
    Department of Physics, Sungkyunkwan University, Suwon 16419, Republic of Korea\label{addr:skku}\and
    Department of Physics and Astronomy, University of Utah, Salt Lake City, UT 84112, USA\label{addr:utah}
}

%% file: source/abstract.tex
\begin{abstract}
We present a comprehensive study of the nonproportionality of NaI(Tl) scintillation detectors within the context of dark matter search experiments.
Our investigation, which integrates COSINE-100 data with supplementary $\gamma$ spectroscopy, measures light yields across diverse energy levels from full-energy $\gamma$ peaks produced by the decays of various isotopes.
These $\gamma$ peaks of interest were produced by decays supported by both long and short-lived isotopes.
Analyzing peaks from decays supported only by short-lived isotopes presented a unique challenge due to their limited statistics and overlapping energies, which was overcome by long-term data collection and a time-dependent analysis.
A key achievement is the direct measurement of the 0.87\,keV light yield, resulting from the cascade following electron capture decay of \iso{Na}{22} from internal contamination.
This measurement, previously accessible only indirectly, deepens our understanding of NaI(Tl) scintillator behavior in the region of interest for dark matter searches.
This study holds substantial implications for background modeling and the interpretation of dark matter signals in NaI(Tl) experiments.
\end{abstract}

%% file: source/introduction.tex
\section{\label{sec:level1}Introduction}

In 1950, Robert W. Pringle observed that the light output of NaI(Tl) scintillation detector is not proportional to the incident photon energy~\cite{pringle_gamma-rays_1950}.
This discovery sparked numerous studies on nonproportionality~(nPR) between light output and the incident electron$/\gamma$ energy using various scintillation detectors~\cite{engelkemeir_nonlinear_1956, aitken_fluorescent_1967, leutz_scintillation_1997, wayne_response_1998, swiderski_response_2013, devare_effect_1963, collinson_fluorescent_1963, jones_nonproportional_1962}.
The nPR phenomenon has been observed in various scintillators, including NaI(Tl), but the detailed values were published differently in each experiment.
To understand the physics behind the phenomenon and the discrepancies observed, numerous remarkable studies have been published over the decades.
These studies include the improvement of measurement techniques~\cite{khodyuk_nonproportional_2010, choong_design_2008, choong_performance_2008, laplace_simultaneous_2021} as well as theoretical interpretations~\cite{moses_origins_2012, payne_nonproportionality_2009, payne_nonproportionality_2011, payne_nonproportionality_2014, payne_nonproportionality_2015, beck_nonproportionality_2015, shi_precise_2002, dietrich_kinetics_1972, rooney_scintillator_1997, bizarri_role_2009, vasilev_luminescence_2008, murray_scintillation_1961, li_transport-based_2011, li_role_2011, kerisit_computer_2009, gao_electron-hole_2008, kerisit_kinetic_2008, bizarri_analytical_2009}, with a subset focusing on simulations.

In recent years, the dark matter~(DM) community has shown increasing interest in the nPR of NaI(Tl), particularly in the low-energy region.
It is driven by experiments searching for DM using NaI(Tl) detectors, inspired by the observations made by DAMA/LIBRA~\revised{\cite{bernabei_dark_2004, bernabei_final_2013, b_bernabei_first_2018, dama_further_2021}}.
DAMA/LIBRA claimed to have detected an annual modulation signal \revised{compatible with DM interaction} below 6\,keV, which has motivated other experiments to analyze low-energy scintillation events in an attempt to replicate their results~\cite{adhikari_lowering_2021, coarasa_improving_2022}.
Moreover, the search for low-mass DM signal in the low-energy region has intensified~\revised{\cite{Workman_2022ynf_dark_matter, billard_direct_2022}}, with consideration of interactions such as weakly interacting massive particle~(WIMP) scattering on the nucleus~\cite{lee_cosmological_1977, goodman_detectability_1985, the_cosine-100_collaboration_experiment_2018, adhikari_strong_2021, adhikari_three-year_2022, amare_annual_2021}, the Migdal effect~\cite{migdal_ionization_1941, ibe_migdal_2018, cosine-100_collaboration_searching_2022}, and DM-electron scattering~\cite{cosine-100_collaboration_search_2023, the_darkside_collaboration_constraints_2018, lee_modulation_2015, griffin_extended_2021, pandax-ii_collaboration_search_2021,essig_direct_2012, graham_semiconductor_2012}.
Another phenomenon that relies on the analysis of low-energy scintillation is coherent elastic neutrino-nucleus scattering~(CE$\nu$NS).
Ongoing efforts aim to observe CE$\nu$NS of reactor neutrinos using NaI(Tl) detectors~\cite{choi_exploring_2023}.
Additionally, a feasibility test of future NaI(Tl) detectors for CE$\nu$NS observations of solar and supernova neutrinos has been conducted~\cite{ko_sensitivities_2023}.
These endeavors crucially require understanding extremely low-energy events.

While substantial efforts have been dedicated to lowering the energy threshold through hardware and software upgrades~\cite{choi_improving_2020, lee_scintillation_2022, adhikari_lowering_2021}, there has been a relative neglect of the study on nPR of scintillation response.
However, misconceptions regarding scintillation response could lead to incorrect interpretations of DM and $\nu$ signals.
\revised{It is also evident in the recent measurements and interpretations of nuclear recoil quenching factors, where taking nPR into account results in an energy scale difference of about 25\% at 1 keV electron equivalent~\cite{lee_quenchingfactor_2024}.}
This is particularly relevant in the low-energy region, where light yield rapidly changes and can introduce significant biases.
Unfortunately, previously measured nPR curves have not adequately covered the low-energy region, resulting in a scarcity of reported light-yield measurements at these energies, with only a few indirect or obsolete values available~\cite{khodyuk_nonproportional_2010, aitken_fluorescent_1967}.

To address the aforementioned challenges, this paper presents the results of our $\gamma$-ray calibration of NaI(Tl) crystals used in the COSINE-100 experiment.
We utilized various mono-energetic peaks from internally contaminated isotopes, covering an energy range from 0.87 to 88\,keV.
Given the long-term nature of the DM search experiment, we extracted these energy peaks by simultaneously examining their decay times and peak positions.
To complement the data, we also conducted $\gamma$~spectroscopy using another NaI(Tl) crystal from the same manufacturer.

In Sec.~\ref{chap:Experiment}, we describe the experimental setup, encompassing the COSINE-100 experiment, as well as the additional $\gamma$~spectroscopy.
COSINE-100 data, which forms the basis of this study, is summarized in Sec.~\ref{chap:InternalPeaks}, highlighting the observations and fitting procedures.
Sec.~\ref{chap:Spectroscopy} focuses on the $\gamma$~spectroscopy methodology, outlining the measurements and analysis techniques to supplement COSINE-100 data.
In Sec.~\ref{chap:Result}, we consolidate the results and present the measured nPR.
Additionally, we provide the resolution and validate it using a waveform simulation.
Finally, in Sec.~\ref{chap:Conclusion}, we conclude by summarizing the key findings of our research and emphasizing the implication for future research.

%% file: source/cosine.tex
\section{\label{chap:Experiment}Experimental Setup}

The COSINE-100 experiment employed eight high-purity NaI(Tl) crystals\revised{ developed in collaboration with Alpha Spectra Inc.
These crystals collectively served as the target for DM direct detection.
They} had a total mass of 106\,kg and were equipped with one 3-inch photomultiplier tube~(PMT) at each of the two ends, enabling the detection of scintillation light.
Each crystal was housed within a copper encapsulation.
The setup featured eight such encapsulations submerged in a 2,200\,L volume of LAB-based liquid scintillator~(LS), which served as an active veto system~\cite{adhikari_cosine-100_2021}.
An acrylic box contained the LS, with eighteen PMTs attached to its walls for scintillation light detection, respectively.
Events were categorized as multiple-hit or single-hit events based on whether particles left signals in other detectors (LS or crystal).
The acrylic box is surrounded by copper and lead, which provide passive shielding layers, while an outermost layer consists of plastic scintillator panels used as a muon veto~\cite{prihtiadi_muon_2018}.

Signals collected by the crystal PMTs were recorded, as waveforms, through a data acquisition~(DAQ) system.
The DAQ system comprised pre-amplifiers, analog-to-digital converters~(ADC), and a trigger control board~(TCB), with detailed specifications outlined in Sec.~7 of Ref.~\cite{adhikari_initial_2018}.
The charge deposition of an event, serving as an indicator of energy, is determined through the integration of the waveform over a 5\,$\mu$s time window following noise removal.
Data taking for physics purposes commenced on October 20, 2016, at the Yangyang Underground Laboratory~\cite{adhikari_initial_2018}, and this study utilized data collected through June 9, 2022.
Notably, three crystals were excluded from the analysis due to their low light yields or high noise levels~\cite{the_cosine-100_collaboration_experiment_2018, adhikari_three-year_2022}.

Despite the high purity of these NaI(Tl) crystals, radioactive decay signals were observed inside them~\cite{adhikari_background_2021}.
Specifically, in the region of interest of this study, three long-lived isotopes --- \iso{Pb}{210}, \iso{Na}{22}, and \iso{K}{40} --- were clearly identified as contamination inside the crystals.
Moreover, several short-lived isotopes\footnote{Precisely to say, short-lived isotopes without long-lived supporting parent isotopes.} were generated by cosmic rays prior to installation, including \iso{Cd}{109}, \iso{Sn}{113}, \iso{I}{125}, \iso{Te}{121m}, and \iso{Te}{127m}~\cite{barbosa_de_souza_study_2020}. 
Signal rates from these isotopes were characterized by the isotopes' decay times and effectively modeled using Monte Carlo simulations.
This paper reanalyzes the energy peaks associated with these isotopes, which serve as calibration points.

Apart from the crystals used in COSINE-100, a smaller-sized NaI(Tl) crystal from the same manufacturer was utilized for a $\gamma$ spectroscopy experiment.
The crystal had a cylindrical shape with a diameter of 3\,inches and a height of 7\,inches.
It was enveloped in 2--3\,mm of PTFE, enclosed by 2\,mm of aluminum, and fitted with a quartz window.
It featured a 3-inch PMT identical to those used in COSINE-100.
The crystal was further shielded using 5-cm-thick lead bricks \revised{and installed at the IBS HQ ground laboratory in Daejeon, Korea}.
The DAQ system for this crystal also included a pre-amplifier, an ADC module, and a TCB, all with the same specifications as those in COSINE-100.
Notably, a more stringent trigger condition was applied to reduce the high trigger rate arising from noise and background events at the low-energy region.
This setup facilitated measurements of energy responses to external $\gamma$ sources, such as \iso{Am}{241}, \iso{Ba}{133}, \iso{Cd}{109}, and \iso{Cs}{137}.
Importantly, the peaks used for analysis remained independent of the trigger condition.

%% file: source/internal.tex
\section{\label{chap:InternalPeaks}Extracting Internal Peaks}

\revised{
From COSINE-100 data, there are peaks resulting from the decay of isotopes contaminated in the crystals, which were used for calibration.
Throughout the entire experimental period, three peaks from long-lived isotopes --- 49\,keV from \iso{Pb}{210}, 0.87\,keV from \iso{Na}{22}, and 3.2\,keV from \iso{K}{40} --- remained consistently visible.
These peaks were extracted by modeling their shapes and the background components around them from the charge distributions stacked over the entire experiment duration.
Short-lived isotopes also produced additional charge peaks that were visible only in the early data.
To extract them, a time-dependent model for charge distributions was developed in accordance with the decay of each isotope.
The following sections outline the formulation and results of the modeling.}

\subsection{\label{sec:LongLivedIsotopes}Long-Lived Isotopes}

The charge distribution of single-hit events in COSINE-100 revealed a distinct peak from \iso{Pb}{210}.
\revised{
It decays to \iso{Bi}{210} through $\beta$ decay, where 84\% of them are excited at 46.5\,keV.
The subsequent deexcitation of this daughter nucleus emits an energy of 46.5\,keV, while the energy of $\beta$ ray is widely distributed, having an average of 4.2\,keV.
Due to the asymmetric shape of the $\beta$ spectrum and the resolution-smearing effect, the peak position becomes approximately 49\,keV.}

\revised{To effectively represent this composition, the} peak is modeled with the function $f_{\mathrm{Pb}}(q)$, defined as
\begin{equation}
\begin{split}
    f_{\mathrm{Pb}}(q) = & \mathrm{Q} \left(q;~ A,~ q_{\mathrm{min}},~ q_{0} \right) \ast \mathrm{N} \left( q;~ \mu=0,~ \sigma^{2} \right) \\
    & + \mathrm{AG} \left(q;~ B,~ \mu,~ \sigma^{2}_{\mathrm{l}},~ \sigma^{2}_{\mathrm{r}} \right) + Cq + D,
    \label{eq:Pb210}
\end{split}
\end{equation}
where $q$ is the integrated charge from the signal waveform, while all the other parameters are free fitting parameters.
In this equation, the first part ($\mathrm{Q} \ast \mathrm{N}$) represents the 49\,keV peak from \iso{Pb}{210}, where $\mathrm{Q}$ is the quadratic distribution, defined as
\begin{equation}
\begin{split}    
    \mathrm{Q} & \left( q; ~ A, ~ q_{\mathrm{min}}, ~ q_{0} \right) \\
    & ~~ \equiv
    \begin{cases}
        A \left( q - q_{0} \right)^{2},  & \text{for } q_{\mathrm{min}} \leq q \leq q_{0} \\
        0,  & \text{otherwise}
    \end{cases}
    ,
\end{split}
\end{equation}
where $A$ is the normalization factor, $q_{0}$ is the vertex position, and $q_\mathrm{min}$ is the minimum range of the distribution.
In the context \revised{of the decay process of \iso{Pb}{210}}, $q_\mathrm{min}$ represents the charge deposition associated with the energy of the $\gamma$ ray, while $q_{0}$ corresponds to the charge deposition at the maximum total energy.
The symbol $\ast$ denotes the convolution operator, and $\mathrm{N}$ is the normal distribution centered at 0 with its standard deviation given by the parameter $\sigma$.
The convolution of this normal distribution mimics the resolution-smearing effect in the charge distribution.

\begin{figure}
    \includegraphics{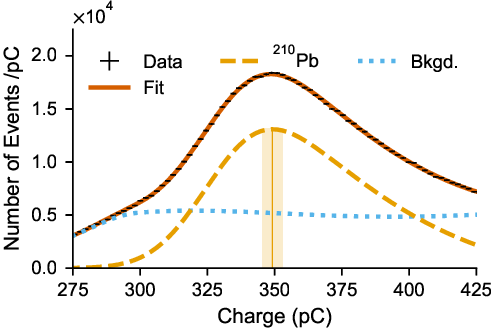}
    \caption{
        \label{fig:Pb210Peak}
        The charge distribution around the 49\,keV peak of \iso{Pb}{210}, as measured by a NaI(Tl) crystal in COSINE-100 (depicted by the black crosses).
        The solid \revised{orange} curve represents the fitted model described in Eq.~\ref{eq:Pb210}.
        The \iso{Pb}{210} peak is drawn as the dashed yellow curve.
        The vertical yellow line indicates the position of the \iso{Pb}{210} peak, along with its associated error.
        The dotted blue curve represents the summed spectrum of background components.
    }
\end{figure}

For the background distribution around this peak, we adopted a sum of a linear component ($Cq + D$) and an asymmetric Gaussian distribution $\mathrm{AG}$ defined as
\begin{equation}
\begin{split}    
    \mathrm{AG} & \left( q;~ B,~ \mu,~ \sigma^{2}_{\mathrm{l}},~ \sigma^{2}_{\mathrm{r}} \right) \\
    & ~ \equiv
    \begin{cases}
        B \exp{\left[ -\left( q - \mu \right)^2 \, / \, 2 \sigma^{2}_{l} \right]}, & \text{for }q < \mu \\
        \\
        B \exp{\left[ -\left( q - \mu \right)^2 \, / \, 2 \sigma^{2}_{r} \right]}, & \text{for }q \geq \mu \\
    \end{cases}
    ,
\end{split}
\end{equation}
where $B$ is the normalization constant, $\mu$ is the peak position, and $\sigma_{\mathrm{l}}$ and $\sigma_{\mathrm{r}}$ are the standard deviations impacting the left and right tails.
\revised{
The distributions accurately represented the background components, as confirmed by the simulated distributions created using the \textsc{Geant4} Monte Carlo simulation toolkit~\cite{agostinelli_geant4simulation_2003}.
The fitting method produced a bias of less than 0.1\% in the peak position, which suggests that it is reliable enough to handle other subdominant radioactive isotopes like \iso{Te}{121m} or \iso{I}{129}.
}
The data distribution and the fitted model are depicted in Fig.~\ref{fig:Pb210Peak}.

The position of the \iso{Pb}{210} peak was determined as the position where the peak component ($\mathrm{Q} \ast \mathrm{N}$) is maximized.
To interpret the position as the light amount for the energy, we assign a 1\% systematic uncertainty to account for the unknown spatial distribution of \iso{Pb}{210} within the crystal and for the poorly known spatial dependence of the detector response.
The total uncertainty is visually depicted as a yellow band around the vertical line representing the peak position in Fig.~\ref{fig:Pb210Peak}.
\revised{
The stability of fitting was improved by assigning the peak position based on its maximum value.
However, this also caused minor fluctuations in the true peak energies among the detectors due to their resolution difference.
The bias term was determined to be about 0.5\% and was considered when calculating the light yield.
}

\begin{figure}
    \includegraphics{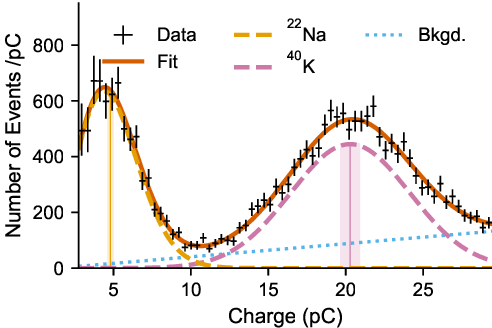}
    \caption{
        \label{fig:LowEnergy}
        The charge distribution of multiple hit events at the low-energy region, as recorded by a NaI(Tl) crystal in COSINE-100 (indicated by the black crosses).
        The solid \revised{orange} curve corresponds to the fitted model.
        Additionally, the dashed yellow curve represents the scaled Poisson distribution for the 0.87\,keV from \iso{Na}{22}, while the dashed pink curve illustrates the normal distribution for the 3.2\,keV from \iso{K}{40}.
        The vertical lines and bands depict their peak positions and associated errors.
        The dotted blue curve represents the linear-shape model for the background components.
    }
\end{figure}

In the charge distribution of multiple-hit events, we observed two distinct peaks associated with isotopes from internal contamination, specifically \iso{Na}{22} and \iso{K}{40}.
Approximately 10\% of \iso{Na}{22} decays to the 1275\,keV level of \iso{Ne}{22} through electron capture~(EC), while a similar proportion of \iso{K}{40} decays via EC to the 1461\,keV level of \iso{Ar}{40}~\cite{BASUNIA201569, CHEN20171}.
The daughter atoms undergo cascade processes, leaving characteristic peaks at their respective K-shell edges within the crystal where \iso{Na}{22} and \iso{K}{40} originally resided.
Concurrently, high-energy $\gamma$ rays are emitted due to the deexcitation of these daughter nuclei.
These energetic $\gamma$ rays can escape the crystal, potentially triggering the neighboring crystals or the LS veto system, leading to coincidences.
Consequently, we clearly identified two low-energy peaks in the charge distribution of multiple-hit events, corresponding to 0.87\,keV and 3.2\,keV.
These values represent the K-shell binding energies of \iso{Ne}{22} and \iso{Ar}{40}~\cite{NISTXray2005}.
The charge distribution is shown in Fig.~\ref{fig:LowEnergy}, where the observed peak positions and shapes align with our expectations.

For the 0.87\,keV peak, a scaled Poisson distribution was used for modeling. The distribution is represented as
\begin{equation}
    f_{\mathrm{Na}} \left( q; ~ A, ~ \lambda, ~ s \right) \equiv
    A\,\frac{\lambda^{q / s }\,e^{-\lambda}}{\Gamma \left( q / s \, + 1 \right)} ,
    \label{eq:ScaledPoisson}
\end{equation}
where $A$ is the normalization constant, $\lambda$ is the expected number of photo-electrons ($N_{\mathrm{pe}}$), and $s$ denotes the scaling factor originating from the PMTs.
The 3.2\,keV peak produced sufficient photo-electrons such that we modeled it using a normal distribution.
The background components were attributed to Compton scattering of high-energy external $\gamma$ rays, resulting in relatively smooth shapes.
To account for the uncertainty in the background distribution's shape, flat or linear-shaped distributions were adopted as background models, and the difference between results from these models was considered as a systematic uncertainty.
All the parameters denoting the normalization, position, and width of the peaks, and the shape of background distribution are free in the fitting process.
The charge distribution, fitted model, and the extracted peak positions are shown in Fig.~\ref{fig:LowEnergy}, and the peak positions are summarized in Table~\ref{tab:InternalPeakPoints}.

\begin{table*}
    \caption{
        \label{tab:InternalPeakPoints}
        List of peak positions extracted from COSINE-100 data.
    }
    \begin{tabular*}{\textwidth}{@{\extracolsep{\fill}}rcccccc@{}}
        \toprule
         & & \multicolumn{5}{c}{Peak Position (pC)} \\
         \cmidrule(lr){3-7}
         Isotope & Energy (keV) & Crystal 2 & Crystal 3 & Crystal 4 & Crystal 6 & Crystal 7 \\
         \midrule[\heavyrulewidth]
         \iso{Na}{22}  & $0.87$ &  $5.29 \, \pm 0.16$ & $5.33 \, \pm 0.17$  & $4.80 \, \pm 0.14$  & $5.00 \, \pm 0.09$  & $5.23 \, \pm 0.11$  \\
         \iso{K}{40}   & $3.2$  & $21.53 \, \pm 0.36$ & $21.70 \, \pm 0.39$ & $20.29 \, \pm 0.58$ & $19.10 \, \pm 0.64$ & $20.61 \, \pm 0.37$ \\
         \iso{Cd}{109} & $25$   & $192.5 \, \pm 2.6$  & $198.3 \, \pm 2.1$  & $183.3 \, \pm 1.9$  & $165.8 \, \pm 2.4$  & $177.5 \, \pm 2.6$  \\
         \iso{Sn}{113} & $28$   &                     & $223.0 \, \pm 3.7$  & $202.3 \, \pm 2.1$  & $190.1 \, \pm 2.2$  & $201.7 \, \pm 2.4$  \\
         \iso{I}{125}  & $39$   &                     & $291.7 \, \pm 5.4$  & $274.1 \, \pm 2.8$  & $265.6 \, \pm 2.8$  & $276.5 \, \pm 2.9$  \\
         \iso{Pb}{210} & $49$   & $370.7 \, \pm 3.7$  & $380.1 \, \pm 3.8$  & $349.2 \, \pm 3.5$  & $337.0 \, \pm 3.4$  & $351.7 \, \pm 3.5$  \\
         \iso{I}{125}  & $67$   &                     & $536.1 \, \pm 7.0$  & $489.0 \, \pm 4.9$  & $467.8 \, \pm 4.7$  & $487.7 \, \pm 4.9$  \\
         \iso{Cd}{109} & $88$   & $674.5 \, \pm 8.4$  & $685.9 \, \pm 7.0$  & $627.2 \, \pm 6.3$  & $605.6 \, \pm 6.2$  & $632.3 \, \pm 6.5$  \\
        \bottomrule
    \end{tabular*}
\end{table*}

\subsection{\label{sec:ShortLivedIsotopes}Short-Lived Isotopes}

Short-lived isotopes, such as \iso{Cd}{109}, \iso{Sn}{113}, \iso{I}{125}, \iso{Te}{121m}, and \iso{Te}{127m}, were initially activated by cosmic rays while the crystals were exposed on the ground and disappeared quickly due to their fast decay times.
Their decay processes emit $\gamma$ rays and X-rays at various energy levels, making them valuable for this study.
Unfortunately, their energy levels overlap, and their activities are scarce, posing challenges for extracting their peaks using the method outlined in Sec.~\ref{sec:LongLivedIsotopes}.
To overcome this issue, we leveraged constraints based on their decay characteristics.

\begin{figure}
    \includegraphics{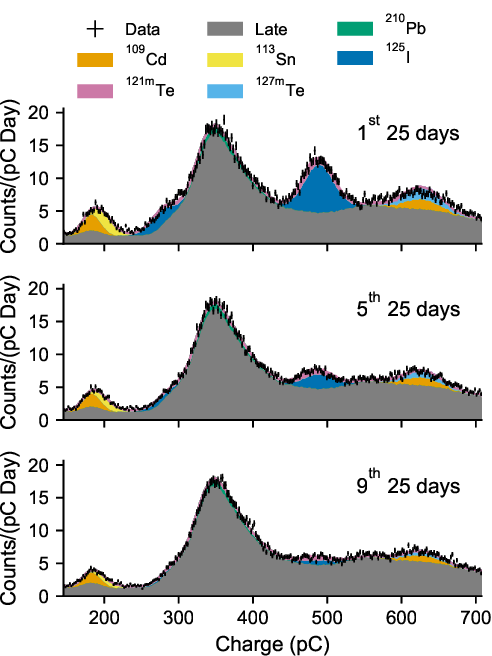}
    \caption{
        \label{fig:Cosmogenic}
        The charge distributions from the initial COSINE-100 data and the extracted peaks attributed to short-lived isotopes.
        The data points, indicated by black crosses, represent measurements from each data collection period and have been simultaneously fitted by using Eq.~\ref{eq:ShortLivedLikelihood}.
        The time-normalized late distribution $L (q)$ is depicted as gray histograms in each data collection period.
        Each isotope component is visualized as a stacked histogram in its own color.
    }
\end{figure}

To observe the decays of these isotopes, we accumulated charge distributions over 475\,days from the beginning of data acquisition, dividing it into 19 periods, each spanning 25\,days.
Fig.~\ref{fig:Cosmogenic} shows the distributions for three periods: $1^\mathrm{st}$, $5^\mathrm{th}$, and $9^\mathrm{th}$.
While decaying and constant components are visually identifiable, determining the position of each peak directly from the spectral shape is challenging.
We built a model to extract the peaks from these spectra, with consideration of the decay characteristics of the short-lived isotopes.

The time-dependent charge distributions for each short-lived isotope $i$ can be expressed as
\begin{equation}
    D_{i} \left(q, ~ t; ~ A_{i}, ~ \vec{\mu}_{i}, ~ \vec{\sigma}_{i} \right) = A_{i} \, e^{-t / \tau_{i}} \sum_{j} f_{ij} \, \mathrm{N} \left( q; ~ \mu_{ij} , ~ \sigma^{2}_{ij} \right),
    \label{eq:PeakComponent}
\end{equation}
where $t$ denotes time, $A_{i}$ and $\tau_{i}$ are the initial activity and lifetime of isotope $i$, and $f_{ij}$ is the branching fraction for the $j$-th peak of isotope $i$.
Each peak is modeled as a normal distribution $\mathrm{N} (q)$ with peak position $\mu_{ij}$ and resolution $\sigma_{ij}$.
For the \iso{Pb}{210} component, the normal distribution in Eq.~\ref{eq:PeakComponent} was replaced with $\mathrm{Q} \ast \mathrm{N}$, as used in Sec.~\ref{sec:LongLivedIsotopes}.

The charge distribution accumulated between times $t_{s}$ and $t_{e}$ can be modeled as
\begin{equation}
\begin{split}
    S & \left( q, ~ t_{s}, ~ t_{e}; ~ \left\{ A_{i}, ~ \vec{\mu}_{i}, ~ \vec{\sigma}_{i} \right\} \right) \\
    & = \int_{t_{s}}^{t_{e}} \left[ \sum_{i} D_{i} \left( q, ~ t; ~ A_{i}, ~ \vec{\mu}_{i}, ~ \vec{\sigma}_{i} \right) + C \left( q \right) \right] s \left( t \right) dt ,
    \label{eq:FullSpectrum}
\end{split}
\end{equation}
where $s(t)$ denotes the detector operational state (1 or 0) at time $t$.
The term $C (q)$ represents the time-independent component in the spectrum, consisting of isotopes with decay times orders of magnitude longer than the experiment's duration.
Instead of constructing a complex analytic model for the uninteresting $C (q)$ term, we attempted to eliminate its explicit dependence.
Since the term is common to all time periods, subtracting a charge distribution from another period with proper time-normalization could effectively remove the $C (q)$ dependency.

For the \revised{subtraction}, a late distribution $L (q)$ was accumulated, expressed as $L \left( q \right) = S \left( q, ~ t_{0}, ~ t_{1} \right)$ for $t_{0}$ and $t_{1}$ defined as 650\,days and 1,150\,days after the start of data acquisition.
Due to its long time period, $L (q)$ has a sufficiently small error to be treated in the model as a known distribution rather than fitted data.
Replacing the $C (q)$ term with $L (q)$ from Eq.~\ref{eq:FullSpectrum} yields
\begin{equation}
\begin{split}
    S & \left( q, ~ t_{s}, ~ t_{e}; ~ \left\{ A_{i}, ~ \vec{\mu}_{i}, ~ \vec{\sigma}_{i} \right\} \right) \\
    & = \int_{t_{s}}^{t_{e}} \sum_{i} D_{i} \left( q, ~ t \right) \, s \left( t \right) \, dt \\
    & ~~~~ + \frac{\int_{t_{s}}^{t_{e}} s(t) \, dt}{\int_{t_{0}}^{t_{1}} s(t) \, dt}
    \left[
        L \left( q \right)
        - \int_{t_{0}}^{t_{1}} \sum_{i} D_{i} \left( q, ~ t \right) \, s \left( t \right) \, dt
    \right]
    ,
    \label{eq:FullSpectrumWithLq}
\end{split}
\end{equation}
which depends only on the set of parameters related to the peaks.
Consequently, the number of events for the $k$-th 25-days-period spectrum in the $l$-th charge bin is expected as $S_{kl} = S \left( q_{l}, ~ (k - 1) \Delta T, ~ k \Delta T \right)$, where $q_{l}$ is the central value of the $l$-th charge bin, and $\Delta T$ is the period interval, 25\,days.

To extract peaks from the short-lived isotopes, we conducted a simultaneous fit on 19 distributions using a binned likelihood,
\begin{equation}
\begin{split}
    \mathcal{L} & \left( \left\{ A_{i}, \vec{\mu}_{i}, ~ \vec{\sigma}_{i} \right\} ; ~ \mathbf{O}, ~ \vec{\mu}^{A}, ~ \vec{\sigma}^{A} \right) \\
    & = \prod_{k}^{19} \prod_{l}^{N_{\mathrm{bin}}} \frac{ S_{kl}^{O_{kl}} e^{-S_{kl}} }{ O_{kl}! } 
    \times \prod_{i} \exp \left[- \frac{1}{2} \left( \frac{A_{i} - \mu_{i}^{A}}{\sigma_{i}^{A}} \right)^{2} \right],
    \label{eq:ShortLivedLikelihood}
\end{split}
\end{equation}
where $O_{kl}$ is the observed number of events in the $k$-th period spectrum at the $l$-th charge bin.
The initial activity $A_{i}$, peak positions $\mu_{ij}$, and resolutions $\sigma_{ij}$ were treated as free parameters, while those for \iso{Pb}{210} were kept constant at the values obtained in Sec.~\ref{sec:LongLivedIsotopes}.
Constraints for the initial activities, $\mu^{A}_{i}$ and $\sigma^{A}_{i}$ were obtained from our previous measurement~\cite{barbosa_de_souza_study_2020}.
As seen in Fig.~\ref{fig:Cosmogenic}, the extracted peaks effectively account for the charge distributions in different time domains simultaneously.

For verification, we applied the same fitting procedure to simulated distributions, allowing us to obtain the true mean energy value for each peak.
Through this process, light yields were obtained for several points ranging from 25 to 88\,keV, as summarized in Table~\ref{tab:InternalPeakPoints}.
\revised{
Crystal 2, cooled underground for about 2.75 years before the experiment started, had the lowest amount of short-lived cosmogenically induced isotopes~\cite{barbosa_de_souza_study_2020}.
It was the longest cooling time among the crystals, resulting in larger errors and missing elements in Table~\ref{tab:InternalPeakPoints} for Crystal 2.
The true energy estimation from the simulated distribution for Crystal 2 was also unstable due to the same reason, and these uncertainties were considered in the nPR analysis.
Moreover, the nPR analysis excluded the peaks from isotopes \iso{Te}{121m} and \iso{Te}{127m} because their amounts were small and their peak positions overlapped with stronger isotopes.
}

%% file: source/spectroscopy.tex
\section{\label{chap:Spectroscopy}Gamma Spectroscopy}

\begin{figure}
    \includegraphics{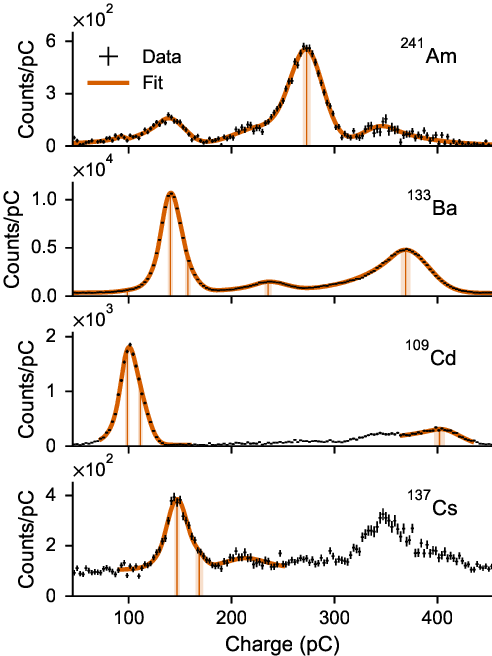}
    \caption{
        \label{fig:Spectroscopy}
        Charge distributions of external sources after subtraction of the background distribution, recorded by the sample crystal for $\gamma$ spectroscopy.
        Black crosses represent the data, and \revised{orange} curves depict the fitting results used to extract the peak positions.
        The peak positions extracted are shown as vertical lines with error bands.
    }
\end{figure}

An independent experiment was conducted for $\gamma$ spectroscopy to acquire complementary data and broaden the energy range of our analysis.
Charge distributions were measured using four $\gamma$ sources, \iso{Am}{241}, \iso{Ba}{133}, \iso{Cd}{109}, and \iso{Cs}{137}.
These sources were placed near the crystal and aligned at its center during data taking.
The energies of the $\gamma$ rays and X-rays emitted from these sources, along with their exposure times, are listed in Table~\ref{tab:SpectroscopyPeakPoints}.
Data without any sources was also collected over an extended period to obtain the background distributions before and after acquiring data with the sources.
This allowed us to verify the stability of the PMT gain and background distribution with time.
The pure peak for each source was identified by subtracting the background distribution, as illustrated in Fig~\ref{fig:Spectroscopy}.

\begin{table}[b]
    \caption{
        \label{tab:SpectroscopyPeakPoints}
        List of sources used for $\gamma$ spectroscopy, with measured peak positions.
        \revised{The energy was obtained by modeling the simulated spectra.}
    }
    \revised{
    \begin{tabular*}{\columnwidth}{@{\extracolsep{\fill}}cccc@{}}
        \toprule
        $\gamma$ Source &   Time  & Energy & Peak Position \\
                & (Hours) &  (keV) & (pC) \\
        \midrule[\heavyrulewidth]
        No Source & 22 & & \\
        \hline
        \iso{Am}{241} & 14 & $59.25 \, \pm 0.16$ & $272.9 \, \pm 2.7$ \\
        \hline
        \multirow{5}{*}{\iso{Ba}{133}} & \multirow{5}{*}{6} & $29.17 \, \pm 0.08$ & $132.6 \, \pm 1.4$ \\
        && $30.82 \, \pm 0.08$ & $140.5 \, \pm 1.4$ \\
        && $35.73 \, \pm 0.10$ & $157.6 \, \pm 2.6$ \\
        && $51.65 \, \pm 0.14$ & $235.5 \, \pm 2.4$ \\
        && $80.00 \, \pm 0.22$ & $369.1 \, \pm 3.7$ \\
        \hline
        \multirow{3}{*}{\iso{Cd}{109}} & \multirow{3}{*}{8} & $21.98 \, \pm 0.06$ & $157.5 \, \pm 1.6$ \\
        && $24.63 \, \pm 0.07$ & $111.5 \, \pm 1.1$ \\
        && $87.54 \, \pm 0.24$ & $402.0 \, \pm 4.3$ \\
        \hline
        \multirow{2}{*}{\iso{Cs}{137}} & \multirow{2}{*}{11} & $32.11 \, \pm 0.10$ & $147.0 \, \pm 1.5$ \\
        && $37.54 \, \pm 0.30$ & $168.6 \, \pm 2.6$ \\
        \bottomrule
    \end{tabular*}
    }
\end{table}

The peak positions were determined by fitting the charge distributions. The charge distribution for isotope $i$ can be modeled as
\begin{equation}
    S_{i} \left( q \right) = \sum_{j} N_{ij} \, \mathrm{CB} \left( q; ~ \mu_{ij}, ~ \sigma_{ij}, ~ \alpha_{ij}, ~ n_{ij} \right),
\end{equation}
where $N_{ij}$ represents the normalization constant of the $j$-th peak, and $\mathrm{CB} (q)$ denotes a Crystal Ball function commonly used to model peaks measured by scintillation crystals with energy loss processes~\cite{gaiser_charmonium_1982}.
The function includes parameters for peak position $\mu_{ij}$ and resolution $\sigma_{ij}$, which are free during the fitting process.
The $\alpha_{ij}$ and $n_{ij}$ parameters account for \revised{the threshold and the exponent of the power-law low-end tail part in the distribution, respectively}.
Flat and linear background components were added to the \iso{Ba}{133} and \iso{Cs}{137} distributions, respectively, to model Compton scattering of high-energy $\gamma$ rays.

The model was fitted to data by minimizing the following chi-squared:
\begin{equation}
    \chi^{2}_{i} \equiv \sum_{l}^{N_{\mathrm{bin}}}
        \dfrac{
            \left[ D_{il} - \left(t_{i}/t_{b}\right) B_{l} - S_{i} \left( q_{l} \right) \right]^{2}
        }{
            D_{il} + \left(t_{i}/t_{b}\right)^{2} B_{l}
        }.
    \label{eq:SpectroscopyChi2}
\end{equation}
In this equation, \revised{$D_{il}$} and $B_{l}$ are the measured number of events in $i$-th source and background data, respectively, at the $l$-th charge bin, centered at $q_{l}$.
Exposure times for the $i$-th source and background data are represented by $t_{i}$ and $t_{b}$, respectively.
The extracted peak positions are displayed in Fig.~\ref{fig:Spectroscopy} as vertical lines with their uncertainties.

\revised{The same fitting procedure was conducted with simulated distributions, as done in Sec.~\ref{chap:InternalPeaks}}.
All dimensions of the detector geometry were implemented based on measurements, except for the thickness of the PTFE wrapped inside the aluminum case.
To account for the uncertainty, two different conditions were simulated with this thicknesses set to 2\,mm and 3\,mm.
The simulation results were fitted separately, and the difference in peak positions was assigned to the systematic error of the true energy.
One can see those errors and the extracted peak positions in Table~\ref{tab:SpectroscopyPeakPoints}.

%% file: source/result.tex
\section{\label{chap:Result}Scintillation Response of NaI(Tl) Crystal}

\revised{
This section presents the nPR of NaI(Tl) crystals based on the measurements described in the previous sections.
We also compare our results with previous references.
In addition, the energy resolution for each peak was measured and validated using a waveform simulation developed, considering the characteristics of the NaI(Tl) scintillator~\cite{choi_waveform_2024}.
}

\subsection{\label{sec:PhotonResponse}Photon Response Measurements}

\begin{figure*}
\begin{center}
    \includegraphics{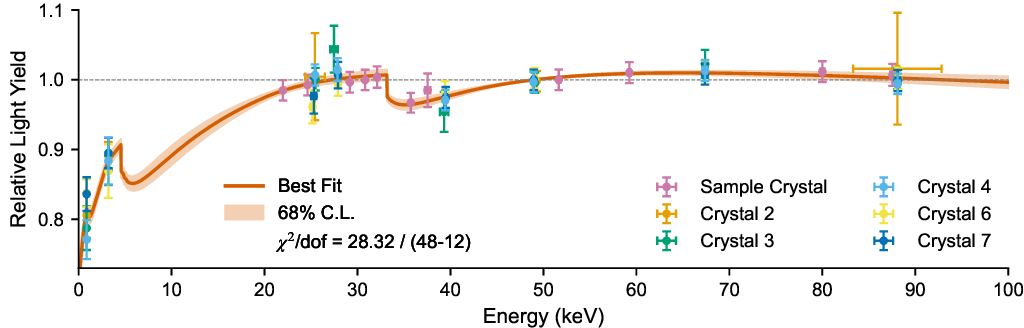}
    \caption{
        \label{fig:Nonproportionality}
        The relative light yield as a function of incident $\gamma$-ray energy, normalized to the value at \revised{49}\,keV.
        The measurements are represented by colored dots with error bars, and the fitted function is drawn as the \revised{solid orange} curve.
        The translucent \revised{orange} band represents the uncertainty from the modeling at \revised{68}\% confidence level.
    }
\end{center}
\end{figure*}

The relative light yield obtained via peak extraction is depicted in Fig.~\ref{fig:Nonproportionality}.
\revised{
In the previous sections, we measured numerous charges of peaks from the data and estimated the true energy values from simulated spectra.
The charge was divided by the true energy for each peak.
To emphasize nonproportional behavior over the absolute values, the ratio of charge to energy was normalized by a constant for each crystal, termed the relative light yield.
Normalization was chosen to set the value to unity at 49\,keV for all crystals.
}
The error bars include systematic errors resulting from the radioactive source's position and the peak extraction method.
These measurements exhibited consistency across the crystals and the measuring techniques within the associated uncertainties.
The well-known K-shell dip of NaI(Tl) at 33\,keV is also identifiable.

The data was fitted using an empirical function developed based on the Payne's model~\cite{payne_nonproportionality_2009, payne_nonproportionality_2011, payne_nonproportionality_2014, payne_nonproportionality_2015, beck_nonproportionality_2015}, as shown by the \revised{orange} curve in Fig.~\ref{fig:Nonproportionality}.
An expectation arises for the presence of the L-shell dip structure around 4.5\,keV, although there is an insufficient number of data points below 20\,keV to constrain the detailed shape of the dip.
We have checked the consistency of our curve with measurements from two references comparing quantities, such as the light-yield ratio between 4.5 and 20\,keV, the ratio between 0.87 and 20\,keV, and the slope around 10\,keV.
These were found to be consistent with previous reports~\cite{khodyuk_nonproportional_2010, aitken_fluorescent_1967}, as summarized in Table~\ref{tab:ReferenceComparison}.
Despite this consistency check, we acknowledge that our curve between 4.5 and 20\,keV is less convincing, resulting in a relatively wider error band, with its width dependent on the selected model parameterization.
The measurement of light yield in this energy region is left for future investigation.

\begin{table}[b]
    \caption{
        \label{tab:ReferenceComparison}
        Comparison of $\gamma$ nPR measurements.
        $\mathrm{LY}_{\gamma} \left( E \right)$ is the light yield for a $\gamma$ ray with incident energy $E$, and \revised{the relative light yield} $\mathrm{LY}_{\gamma,\,\revised{49}} \left( E \right) \equiv \mathrm{LY}_{\gamma} \left( E \right) / \mathrm{LY}_{\gamma} \left( \revised{49}\,\mathrm{keV} \right)$.
    }
    \begin{tabular*}{\columnwidth}{@{\extracolsep{\fill}}cccc@{}}
        \toprule
        & This Work & Khodyuk & Aitken \\
        &  & \revised{\cite{khodyuk_nonproportional_2010}} & \revised{\cite{aitken_fluorescent_1967}} \\
        \midrule
        $\dfrac{d \, \mathrm{LY}_{\gamma,\,\revised{49}}}{dE} \left( 10\,\mathrm{keV} \right)$ $\left( \dfrac{\%}{\mathrm{keV}} \right)$&
            1.12 & 1.18 & 1.16 \\
        \\
        $\dfrac{\mathrm{LY}_{\gamma} \left( 4.5 \, \mathrm{keV} \right)}{\mathrm{LY}_{\gamma} \left( 20 \, \mathrm{keV} \right)}$ (\%)&
            93.0 & 93.3 & 91.8 \\
        \\
        $\dfrac{\mathrm{LY}_{\gamma} \left( \revised{0.87} \, \mathrm{keV} \right)}{\mathrm{LY}_{\gamma} \left( 20 \, \mathrm{keV} \right)}$ (\%)&
            83.7 & 89.3 & \\
        \bottomrule
    \end{tabular*}
\end{table}

\begin{figure}
    \includegraphics{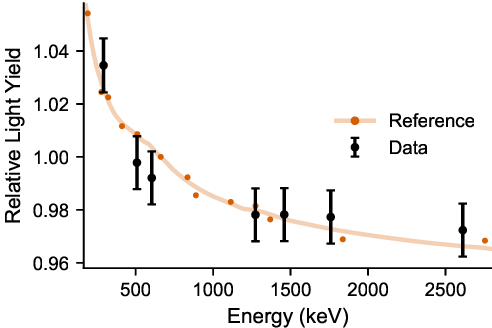}
    \caption{
        \label{fig:DynodeConsistency}
        The relative light yield (normalized at 662\,keV) for the high-energy $\gamma$ peaks is represented by black dots with error bars.
        The \revised{orange dots and solid orange curve} represent the nPR measured from Ref.~\cite{devare_effect_1963}, where our data agrees well with the reference.
    }
\end{figure}

To evaluate the nPR in the high-energy region above 200\,keV, we examined its consistency with the data reported in Ref.~\cite{devare_effect_1963}.
We measured the light yields by fitting the prominent $\gamma$ peaks from the high-energy spectrum, including 295\,keV from \iso{Pb}{214}, 511 and 1274\,keV from \iso{Na}{22}, 609 and 1764\,keV from \iso{Bi}{214}, 1461\,keV from \iso{K}{40}, and 2615\,keV from \iso{Tl}{208}.
The measured light yields demonstrate agreement with the nPR curve reported in Ref.~\cite{devare_effect_1963}, as illustrated in Fig.~\ref{fig:DynodeConsistency}.

\subsection{\label{sec:Resolution}Resolution and Waveform Simulation}

\begin{figure}
    \includegraphics{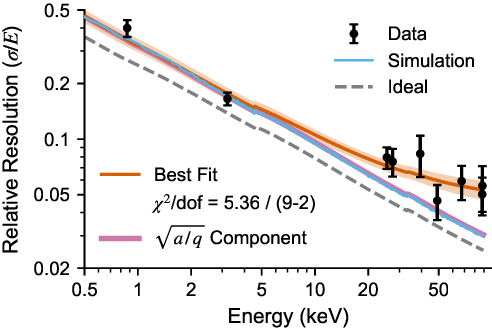}
    \caption{
        \label{fig:Resolution}
        The relative resolution as a function of incident $\gamma$-ray energy for a NaI(Tl) crystal.
        The black dots with error bars represent the measurements, and the \revised{solid orange} curve represents the model of the resolution curve, as described in Eq.~\ref{eq:Resolution}, with an error band at \revised{68}\% confidence level.
        The solid pink curve illustrates the $\sqrt{a/q}$ component from the resolution model, which closely matches the waveform simulation (solid blue curve).
        The dashed gray curve represents the Poisson fluctuation in $N_{\mathrm{pe}}$.
    }
\end{figure}

Once a peak is extracted, the energy resolution $\sigma_{ij}$ is naturally obtained along with the peak position $\mu_{ij}$.
As shown in the dashed gray curve in Fig.~\ref{fig:Resolution}, the characteristic $1/\sqrt{E}$-like shape, originating from the Poisson fluctuations in $N_{\mathrm{pe}}$ for the PMT, is clearly visible.
Also resolution effects due to the detector geometry and the distribution of the number of amplified electrons in the PMT are observed.

To account for these effects, the relative resolution as a function of charge was modeled using the following equation,
\begin{equation}
    \frac{\sigma (q)}{q} = \sqrt{\frac{a}{q} + b}.
    \label{eq:Resolution}
\end{equation}
Here, $a$ and $b$ are fitting parameters, where $a$ scales the Poisson resolution contribution and $b$ typically arises from non-uniformity of detector response, which is not included in the simulation.
In Fig.~\ref{fig:Resolution}, the \revised{orange} line represents the model fitted to the measurements, with the uncertainty shown as a band.
The resolution $\sigma (q)$ is transformed to $\sigma (E)$ via the nPR curve shown in Fig.~\ref{fig:Nonproportionality}, leading to the emergence of non-analytic points at the 4\,keV L-shell edge and the 33\,keV K-shell edge.
The pink line represents the function when $b = 0$.

Given the focus on DM detection experiments, where the region of interest typically lies below 6\,keV, it is crucial to investigate the resolution in the sub-keV range, particularly for future upgrades.
Since the effect of electron amplification in PMTs and DAQ system is most dominant at low energies, except for that due to the Poisson fluctuations with $N_{\mathrm{pe}}$, a waveform simulation developed for the COSINE-100 experiment was utilized to validate the resolution in this region~\cite{choi_waveform_2024}.
This simulation incorporates factors such as rise and decay times of NaI(Tl) scintillation, fluctuations in the single photo-electron shape, DAQ electronics, and charge accumulation processes.
Therefore, by inputting $N_{\mathrm{pe}}$ that takes into account Poisson fluctuations, we can estimate the resolution with these effects included.

The blue line in Fig.~\ref{fig:Resolution} shows that the simulation successfully replicates the data.
Despite not accounting for the intricate details of detector geometry and other mechanisms, the simulation exhibits good agreement with the measurements at the low-energy region.
In addition, the agreement between the simulation and the pink curve, which is the measurement with the constant-term factor excluded, confirms the possibility of extrapolating the resolution curve to the low-energy direction.
These findings hold particular significance for scintillator experiments focusing on low-energy ranges.

%% file: source/conclusion.tex
\section{\label{chap:Conclusion}Summary and Conclusion}

In conclusion, this study has thoroughly examined the nPR characteristics of NaI(Tl) scintillators within the context of DM search experiments.
Through a comprehensive calibration process using COSINE-100 data, complemented by $\gamma$ spectroscopy, we have gained valuable insights into the behavior of NaI(Tl) crystals.

The measurement of light yields at various energies was successfully achieved by analyzing internal peaks originating from both long-lived and short-lived isotopes.
Extraction of peaks from long-lived isotopes was easily accomplished due to stable experimental data acquisition.
A significant milestone was reached by directly measuring the light yield at 0.87\,keV, generated by the cascade following \iso{Na}{22} EC decay, facilitated by the active veto system.
This direct measurement serves as validation of an indirect measurement obtained by analyzing X-rays near the Iodine K-shell using synchrotron radiation~\cite{khodyuk_nonproportional_2010}.
Overcoming the challenge of extracting multiple overlaid peaks from short-lived isotopes was made possible through long-term data collection and time-dependent analysis.

In the nPR curve measured here, the distinctive K-shell dip structure was observed. The nPR curve was fitted using an empirical curve.
The resulting fitted nPR curve exhibits consistency with previously reported measurements, particularly in the energy range below 20\,keV.
However, the lack of data points around the 10\,keV region limits the precision of the calibration curve, leaving the error band relatively wider.
This aspect remains a focus for future investigation.

The resolution analysis of NaI(Tl) scintillators revealed a characteristic resolution curve that follows the $1/\sqrt{N_{\mathrm{pe}}}$ trend, as expected for Poisson fluctuations in $N_{\mathrm{pe}}$.
Waveform simulation provided compelling confirmation of the resolution at low energies, allowing for the extension of the resolution curve to lower energy ranges.
This enhancement in our ability to detect low-energy DM recoil signals is very important.

The findings of this study hold significant implications for background modeling in NaI(Tl) experiments.
Precise calibration and characterization of the nPR effect are essential for accurately modeling the background spectra resulting from radioactive interactions involving electrons and photons.
To facilitate more detailed simulations, the integration of the nonproportional light production of NaI(Tl) into the \textsc{Geant4} toolkit is under development, building on similar efforts reported previously~\cite{rasco_nonlinear_2015, cano-ott_monte_1999}.

Furthermore, it is imperative to consider nPR calibration when interpreting signals that may come from DM interactions with electrons or photons.
As the scientific community explores diverse scenarios involving boosted DM and light DM, the significance of these interactions is poised to grow.
This study lays the foundation for more robust and accurate DM search experiments in the future.

%% file: source/acknowledgments_September2023_v3.tex
We thank the Korea Hydro and Nuclear Power (KHNP) Company for providing underground laboratory space at Yangyang and the IBS Research Solution Center (RSC) for providing high-performance computing resources. 
This work is supported by the Institute for Basic Science (IBS) under project code IBS-R016-A1, NRF-2019R1C1C1005073, NRF-2021R1A2C3010989, NRF-2021R1A2C1013761 and Hyundai Motor Chung Mong-Koo Foundation, Republic of Korea;
NSF Grants No. PHY-1913742, DGE-1122492, WIPAC, the Wisconsin Alumni Research Foundation, United States; 
STFC Grant ST/N000277/1 and ST/K001337/1, United Kingdom;
Grant No. 2021/06743-1, 2022/12002-7 and 2022/13293-5 FAPESP, CAPES Finance Code 001, CNPq 303122/2020-0, Brazil.

%% file: main.bbl
\providecommand{\href}[2]{#2}\begingroup\raggedright\begin{thebibliography}{10}

\bibitem{pringle_gamma-rays_1950}
R.~W. Pringle and S.~Standil,
  \href{https://doi.org/10.1103/PhysRev.80.762}{{{The {Gamma}-{Rays} from
  {Neutron}-{Activated} {Gold}}}}, Phys. Rev. {\bf 80}, 762--763 (1950).

\bibitem{engelkemeir_nonlinear_1956}
D.~Engelkemeir, \href{https://doi.org/10.1063/1.1715643}{{{Nonlinear {Response}
  of {NaI}({Tl}) to {Photons}}}}, Rev. Sci. Instrum. {\bf 27}, 589--591 (1956).

\bibitem{aitken_fluorescent_1967}
D.~W. Aitken et~al., \href{https://doi.org/10.1109/TNS.1967.4324457}{{{The
  {Fluorescent} {Response} of {NaI}({Tl}), {CsI}({Tl}), {CsI}({Na}) and
  {CaF2}({Eu}) to {X}-{Rays} and {Low} {Energy} {Gamma} {Rays}}}}, IEEE Trans.
  Nucl. Sci. {\bf 14}, 468--477 (1967).

\bibitem{leutz_scintillation_1997}
H.~Leutz and C.~D'Ambrosio, \href{https://doi.org/10.1109/23.568804}{{{On the
  scintillation response of {NaI}({TI})-crystals}}}, IEEE Trans. Nucl. Sci.
  {\bf 44}, 190--193 (1997).

\bibitem{wayne_response_1998}
L.~R. Wayne et~al.,
  \href{https://doi.org/10.1016/S0168-9002(98)00193-4}{{{Response of
  {NaI}({Tl}) to {X}-rays and electrons}}}, Nucl. Instrum. Methods Phys. Res.
  Sect. A {\bf 411}, 351--364 (1998).

\bibitem{swiderski_response_2013}
L.~Swiderski et~al.,
  \href{https://doi.org/10.1016/j.nima.2012.11.188}{{{Response of doped alkali
  iodides measured with gamma-ray absorption and {Compton} electrons}}}, Nucl.
  Instrum. Methods Phys. Res. Sect. A {\bf 705}, 42--46 (2013).

\bibitem{devare_effect_1963}
H.~G. Devare and P.~N. Tandon,
  \href{https://doi.org/10.1016/0029-554X(63)90252-0}{{{Effect of the
  non-linear response of {NaI}({Tl}) on the single crystal summing spectra}}},
  Nucl. Instrum. Methods {\bf 22}, 253--255 (1963).

\bibitem{collinson_fluorescent_1963}
A.~J.~L. Collinson and R.~Hill,
  \href{https://doi.org/10.1088/0370-1328/81/5/313}{{{The {Fluorescent}
  {Response} of {NaI}({Tl}) and {CsI}({Tl}) to {X} {Rays} and γ {Rays}}}},
  Proc. Phys. Soc. {\bf 81}, 883 (1963).

\bibitem{jones_nonproportional_1962}
T.~H. Jones, \href{https://doi.org/10.1016/0029-554X(62)90026-5}{{{The
  nonproportional response of a {NaI}({Tl}) crystal to diffracted {X} rays}}},
  Nucl. Instrum. Methods {\bf 15}, 55--58 (1962).

\bibitem{khodyuk_nonproportional_2010}
I.~V. Khodyuk, P.~A. Rodnyi, and P.~Dorenbos,
  \href{https://doi.org/10.1063/1.3431009}{{{Nonproportional scintillation
  response of {NaI}:{Tl} to low energy x-ray photons and electrons}}}, J. Appl.
  Phys. {\bf 107}, 113513 (2010).

\bibitem{choong_design_2008}
W.-S. Choong et~al., \href{https://doi.org/10.1109/TNS.2008.921491}{{{Design of
  a {Facility} for {Measuring} {Scintillator} {Non}-{Proportionality}}}}, IEEE
  Trans. Nucl. Sci. {\bf 55}, 1753--1758 (2008).

\bibitem{choong_performance_2008}
W.-S. Choong et~al.,
  \href{https://doi.org/10.1109/TNS.2008.922824}{{{Performance of a {Facility}
  for {Measuring} {Scintillator} {Non}-{Proportionality}}}}, IEEE Trans. Nucl.
  Sci. {\bf 55}, 1073--1078 (2008).

\bibitem{laplace_simultaneous_2021}
T.~A. Laplace et~al.,
  \href{https://doi.org/10.1103/PhysRevC.104.014609}{{{Simultaneous measurement
  of organic scintillator response to carbon and proton recoils}}}, Phys. Rev.
  C {\bf 104}, 014609 (2021).

\bibitem{moses_origins_2012}
W.~W. Moses et~al., \href{https://doi.org/10.1109/TNS.2012.2186463}{{{The
  {Origins} of {Scintillator} {Non}-{Proportionality}}}}, IEEE Trans. Nucl.
  Sci. {\bf 59}, 2038--2044 (2012).

\bibitem{payne_nonproportionality_2009}
S.~A. Payne et~al.,
  \href{https://doi.org/10.1109/TNS.2009.2023657}{{{Nonproportionality of
  {Scintillator} {Detectors}: {Theory} and {Experiment}}}}, IEEE Trans. Nucl.
  Sci. {\bf 56}, 2506--2512 (2009).

\bibitem{payne_nonproportionality_2011}
S.~A. Payne et~al.,
  \href{https://doi.org/10.1109/TNS.2011.2167687}{{{Nonproportionality of
  {Scintillator} {Detectors}: {Theory} and {Experiment}. {II}}}}, IEEE Trans.
  Nucl. Sci. {\bf 58}, 3392--3402 (2011).

\bibitem{payne_nonproportionality_2014}
S.~A. Payne et~al.,
  \href{https://doi.org/10.1109/TNS.2014.2343572}{{{Nonproportionality of
  {Scintillator} {Detectors}. {III}. {Temperature} {Dependence} {Studies}}}},
  IEEE Trans. Nucl. Sci. {\bf 61}, 2771--2777 (2014).

\bibitem{payne_nonproportionality_2015}
S.~A. Payne,
  \href{https://doi.org/10.1109/TNS.2014.2387256}{{{Nonproportionality of
  {Scintillator} {Detectors}. {IV}. {Resolution} {Contribution} from
  {Delta}-{Rays}}}}, IEEE Trans. Nucl. Sci. {\bf 62}, 372--380 (2015).

\bibitem{beck_nonproportionality_2015}
P.~R. Beck et~al.,
  \href{https://doi.org/10.1109/TNS.2015.2414357}{{{Nonproportionality of
  {Scintillator} {Detectors}. {V}. {Comparing} the {Gamma} and {Electron}
  {Response}}}}, IEEE Trans. Nucl. Sci. {\bf 62}, 1429--1436 (2015).

\bibitem{shi_precise_2002}
H.-X. Shi et~al.,
  \href{https://doi.org/10.1016/S0969-8043(02)00140-9}{{{Precise {Monte}
  {Carlo} simulation of gamma-ray response functions for an {NaI}({Tl})
  detector}}}, Appl. Radiat. Isot. {\bf 57}, 517--524 (2002).

\bibitem{dietrich_kinetics_1972}
H.~B. Dietrich and R.~B. Murray,
  \href{https://doi.org/10.1016/0022-2313(72)90039-7}{{{Kinetics of the
  diffusion of self-trapped holes in alkali halide scintillators}}}, J.
  Luminescence {\bf 5}, 155--170 (1972).

\bibitem{rooney_scintillator_1997}
B.~Rooney and J.~Valentine,
  \href{https://doi.org/10.1109/23.603702}{{{Scintillator light yield
  nonproportionality: calculating photon response using measured electron
  response}}}, IEEE Trans. Nucl. Sci. {\bf 44}, 509--516 (1997).

\bibitem{bizarri_role_2009}
G.~Bizarri et~al., \href{https://doi.org/10.1016/j.jlumin.2008.12.024}{{{The
  role of different linear and non-linear channels of relaxation in
  scintillator non-proportionality}}}, J. Luminescence {\bf 129}, 1790--1793
  (2009).

\bibitem{vasilev_luminescence_2008}
A.~N. Vasil'ev, \href{https://doi.org/10.1109/TNS.2007.914367}{{{From
  {Luminescence} {Non}-{Linearity} to {Scintillation}
  {Non}-{Proportionality}}}}, IEEE Trans. Nucl. Sci. {\bf 55}, 1054--1061
  (2008).

\bibitem{murray_scintillation_1961}
R.~B. Murray and A.~Meyer,
  \href{https://doi.org/10.1103/PhysRev.122.815}{{{Scintillation {Response} of
  {Activated} {Inorganic} {Crystals} to {Various} {Charged} {Particles}}}},
  Phys. Rev. {\bf 122}, 815--826 (1961).

\bibitem{li_transport-based_2011}
Q.~Li et~al., \href{https://doi.org/10.1063/1.3600070}{{{A transport-based
  model of material trends in nonproportionality of scintillators}}}, J. Appl.
  Phys. {\bf 109}, 123716 (2011).

\bibitem{li_role_2011}
Q.~Li et~al., \href{https://doi.org/10.1016/j.nima.2010.07.074}{{{The role of
  hole mobility in scintillator proportionality}}}, Nucl. Instrum. Methods
  Phys. Res. Sect. A {\bf 652}, 288--291 (2011).

\bibitem{kerisit_computer_2009}
S.~Kerisit et~al., \href{https://doi.org/10.1063/1.3143786}{{{Computer
  simulation of the light yield nonlinearity of inorganic scintillators}}}, J.
  Appl. Phys. {\bf 105}, 114915 (2009).

\bibitem{gao_electron-hole_2008}
F.~Gao et~al., \href{https://doi.org/10.1109/TNS.2007.908917}{{{Electron-{Hole}
  {Pairs} {Created} by {Photons} and {Intrinsic} {Properties} in {Detector}
  {Materials}}}}, IEEE Trans. Nucl. Sci. {\bf 55}, 1079--1085 (2008).

\bibitem{kerisit_kinetic_2008}
S.~Kerisit, K.~M. Rosso, and B.~D. Cannon,
  \href{https://doi.org/10.1109/TNS.2008.922830}{{{Kinetic {Monte} {Carlo}
  {Model} of {Scintillation} {Mechanisms} in {CsI} and {CsI}({Tl})}}}, IEEE
  Trans. Nucl. Sci. {\bf 55}, 1251--1258 (2008).

\bibitem{bizarri_analytical_2009}
G.~Bizarri et~al., \href{https://doi.org/10.1063/1.3081651}{{{An analytical
  model of nonproportional scintillator light yield in terms of recombination
  rates}}}, J. Appl. Phys. {\bf 105}, 044507 (2009).

\bibitem{bernabei_dark_2004}
R.~Bernabei et~al., ({DAMA/NaI}),
  \href{https://doi.org/10.1142/S0218271804006619}{{{{Dark} {Matter}
  {Particles} in the {Galactic} {Halo}: {Results} and {Implications} from
  dama/nai}}}, Int. J. Mod. Phys. D {\bf 13}, 2127--2159 (2004).

\bibitem{bernabei_final_2013}
R.~Bernabei et~al., ({DAMA/LIBRA}),
  \href{https://doi.org/10.1140/epjc/s10052-013-2648-7}{{{Final model
  independent result of {DAMA}/{LIBRA}–phase1}}}, Eur. Phys. J. C {\bf 73},
  2648 (2013).

\bibitem{b_bernabei_first_2018}
B.~Bernabei et~al., ({DAMA/LIBRA}),
  \href{https://doi.org/10.15407/jnpae2018.04.307}{{{First model independent
  results from {DAMA}/{LIBRA}-phase2}}}, Nucl. Phys. At. Energy {\bf 19},
  307--325 (2018).

\bibitem{dama_further_2021}
R.~Bernabei et~al., ({DAMA/LIBRA}),
  \href{https://doi.org/10.15407/jnpae2021.04.329}{{{Further results from
  {DAMA}/{LIBRA}-phase2 and perspectives}}}, Nucl. Phys. At. Energy {\bf 22},
  329--342 (2021).

\bibitem{adhikari_lowering_2021}
G.~Adhikari et~al., ({COSINE-100}),
  \href{https://doi.org/10.1016/j.astropartphys.2021.102581}{{{Lowering the
  energy threshold in {COSINE}-100 dark matter searches}}}, Astropart. Phys.
  {\bf 130}, 102581 (2021).

\bibitem{coarasa_improving_2022}
I.~Coarasa et~al., ({ANAIS-112}),
  \href{https://doi.org/10.1088/1475-7516/2022/11/048}{{{Improving {ANAIS}-112
  sensitivity to {DAMA}/{LIBRA} signal with machine learning techniques}}}, J.
  Cosmol. Astropart. Phys. {\bf 2022}, 048 (2022).

\bibitem{Workman_2022ynf_dark_matter}
R.~L. Workman and Others, ({Particle Data Group}),
  \href{https://doi.org/10.1093/ptep/ptac097}{{{{Review of Particle
  Physics}}}}, PTEP {\bf 2022}, 083C01 (2022). See chapter 27 (Dark matter).

\bibitem{billard_direct_2022}
J.~Billard et~al., \href{https://doi.org/10.1088/1361-6633/ac5754}{{{Direct
  detection of dark matter—{APPEC} committee report}}}, Rep. Prog. Phys. {\bf
  85}, 056201 (2022).

\bibitem{lee_cosmological_1977}
B.~W. Lee and S.~Weinberg,
  \href{https://doi.org/10.1103/PhysRevLett.39.165}{{{Cosmological {Lower}
  {Bound} on {Heavy}-{Neutrino} {Masses}}}}, Phys. Rev. Lett. {\bf 39},
  165--168 (1977).

\bibitem{goodman_detectability_1985}
M.~W. Goodman and E.~Witten,
  \href{https://doi.org/10.1103/PhysRevD.31.3059}{{{Detectability of certain
  dark-matter candidates}}}, Phys. Rev. D {\bf 31}, 3059--3063 (1985).

\bibitem{the_cosine-100_collaboration_experiment_2018}
G.~Adhikari et~al., ({COSINE-100}),
  \href{https://doi.org/10.1038/s41586-018-0739-1}{{{An experiment to search
  for dark-matter interactions using sodium iodide detectors}}}, Nature {\bf
  564}, 83--86 (2018).

\bibitem{adhikari_strong_2021}
G.~Adhikari et~al., ({COSINE-100}),
  \href{https://doi.org/10.1126/sciadv.abk2699}{{{Strong constraints from
  {COSINE}-100 on the {DAMA} dark matter results using the same sodium iodide
  target}}}, Sci. Adv. {\bf 7}, eabk2699 (2021).

\bibitem{adhikari_three-year_2022}
G.~Adhikari et~al., ({COSINE-100}),
  \href{https://doi.org/10.1103/PhysRevD.106.052005}{{{Three-year annual
  modulation search with {COSINE}-100}}}, Phys. Rev. D {\bf 106}, 052005
  (2022).

\bibitem{amare_annual_2021}
J.~Amaré et~al., ({ANAIS-112}),
  \href{https://doi.org/10.1103/PhysRevD.103.102005}{{{Annual modulation
  results from three-year exposure of {ANAIS}-112}}}, Phys. Rev. D {\bf 103},
  102005 (2021).

\bibitem{migdal_ionization_1941}
A.~Migdal,
  \href{https://scholar.google.com/citations?view_op=view_citation&hl=en&user=Tjnr6kgAAAAJ&citation_for_view=Tjnr6kgAAAAJ:txeM2kYbVNMC}{{{Ionization
  of atoms accompanying α-and β-decay}}}, J. Phys. USSR {\bf 4}, 449 (1941).

\bibitem{ibe_migdal_2018}
M.~Ibe et~al., \href{https://doi.org/10.1007/JHEP03(2018)194}{{{Migdal effect
  in dark matter direct detection experiments}}}, J. High Energ. Phys. {\bf
  2018}, 194 (2018).

\bibitem{cosine-100_collaboration_searching_2022}
G.~Adhikari et~al., ({COSINE-100}),
  \href{https://doi.org/10.1103/PhysRevD.105.042006}{{{Searching for low-mass
  dark matter via the {Migdal} effect in {COSINE}-100}}}, Phys. Rev. D {\bf
  105}, 042006 (2022).

\bibitem{cosine-100_collaboration_search_2023}
G.~Adhikari et~al., ({COSINE-100}),
  \href{https://doi.org/10.1103/PhysRevD.107.122004}{{{Search for solar bosonic
  dark matter annual modulation with {COSINE}-100}}}, Phys. Rev. D {\bf 107},
  122004 (2023).

\bibitem{the_darkside_collaboration_constraints_2018}
P.~Agnes et~al., ({DarkSide}),
  \href{https://doi.org/10.1103/PhysRevLett.121.111303}{{{Constraints on
  {Sub}-{GeV} {Dark}-{Matter}--{Electron} {Scattering} from the {DarkSide}-50
  {Experiment}}}}, Phys. Rev. Lett. {\bf 121}, 111303 (2018).

\bibitem{lee_modulation_2015}
S.~K. Lee et~al.,
  \href{https://doi.org/10.1103/PhysRevD.92.083517}{{{Modulation effects in
  dark matter-electron scattering experiments}}}, Phys. Rev. D {\bf 92}, 083517
  (2015).

\bibitem{griffin_extended_2021}
S.~M. Griffin et~al.,
  \href{https://doi.org/10.1103/PhysRevD.104.095015}{{{Extended calculation of
  dark matter-electron scattering in crystal targets}}}, Phys. Rev. D {\bf
  104}, 095015 (2021).

\bibitem{pandax-ii_collaboration_search_2021}
C.~Cheng et~al., ({PandaX-II}),
  \href{https://doi.org/10.1103/PhysRevLett.126.211803}{{{Search for {Light}
  {Dark} {Matter}--{Electron} {Scattering} in the {PandaX}-{II}
  {Experiment}}}}, Phys. Rev. Lett. {\bf 126}, 211803 (2021).

\bibitem{essig_direct_2012}
R.~Essig, J.~Mardon, and T.~Volansky,
  \href{https://doi.org/10.1103/PhysRevD.85.076007}{{{Direct detection of
  sub-{GeV} dark matter}}}, Phys. Rev. D {\bf 85}, 076007 (2012).

\bibitem{graham_semiconductor_2012}
P.~W. Graham et~al.,
  \href{https://doi.org/10.1016/j.dark.2012.09.001}{{{Semiconductor probes of
  light dark matter}}}, Phys. Dark Universe {\bf 1}, 32--49 (2012).

\bibitem{choi_exploring_2023}
J.~J. Choi et~al., ({NEON}),
  \href{https://doi.org/10.1140/epjc/s10052-023-11352-x}{{{Exploring coherent
  elastic neutrino-nucleus scattering using reactor electron antineutrinos in
  the {NEON} experiment}}}, Eur. Phys. J. C {\bf 83}, 226 (2023).

\bibitem{ko_sensitivities_2023}
Y.~J. Ko and H.~S. Lee,
  \href{https://doi.org/10.1016/j.astropartphys.2023.102890}{{{Sensitivities
  for coherent elastic scattering of solar and supernova neutrinos with future
  {NaI}({Tl}) dark matter search detectors of {COSINE}-200/{1T}}}}, Astropart.
  Phys. {\bf 153}, 102890 (2023).

\bibitem{choi_improving_2020}
J.~J. Choi et~al., ({NEON}),
  \href{https://doi.org/10.1016/j.nima.2020.164556}{{{Improving the light
  collection using a new {NaI}({Tl}) crystal encapsulation}}}, Nucl. Instrum.
  Methods Phys. Res. Sect. A {\bf 981}, 164556 (2020).

\bibitem{lee_scintillation_2022}
H.~Y. Lee et~al.,
  \href{https://doi.org/10.1088/1748-0221/17/02/P02027}{{{Scintillation
  characteristics of a {NaI}({Tl}) crystal at low-temperature with silicon
  photomultiplier}}}, J. Inst. {\bf 17}, P02027 (2022).

\bibitem{lee_quenchingfactor_2024}
S.~H. Lee et~al.,
  \href{https://doi.org/10.48550/arXiv.2402.15122}{{{Measurements of low energy
  nuclear recoil quenching factors for {Na} and {I} recoils in the {NaI}({Tl})
  scintillator}}}, \href{http://arxiv.org/abs/2402.15122}{{\tt
  arXiv:2402.15122}} (2024).

\bibitem{adhikari_cosine-100_2021}
G.~Adhikari et~al., ({COSINE-100}),
  \href{https://doi.org/10.1016/j.nima.2021.165431}{{{The {COSINE}-100 liquid
  scintillator veto system}}}, Nucl. Instrum. Methods Phys. Res. Sect. A {\bf
  1006}, 165431 (2021).

\bibitem{prihtiadi_muon_2018}
H.~Prihtiadi et~al., ({COSINE-100}),
  \href{https://doi.org/10.1088/1748-0221/13/02/T02007}{{{Muon detector for the
  {COSINE}-100 experiment}}}, J. Inst. {\bf 13}, T02007 (2018).

\bibitem{adhikari_initial_2018}
G.~Adhikari et~al., ({COSINE-100}),
  \href{https://doi.org/10.1140/epjc/s10052-018-5590-x}{{{Initial performance
  of the {COSINE}-100 experiment}}}, Eur. Phys. J. C {\bf 78}, 107 (2018).

\bibitem{adhikari_background_2021}
G.~Adhikari et~al., ({COSINE-100}),
  \href{https://doi.org/10.1140/epjc/s10052-021-09564-0}{{{Background modeling
  for dark matter search with 1.7 years of {COSINE}-100 data}}}, Eur. Phys. J.
  C {\bf 81}, 837 (2021).

\bibitem{barbosa_de_souza_study_2020}
E.~Barbosa De~Souza et~al., ({COSINE-100}),
  \href{https://doi.org/10.1016/j.astropartphys.2019.102390}{{{Study of
  cosmogenic radionuclides in the {COSINE}-100 {NaI}({Tl}) detectors}}},
  Astropart. Phys. {\bf 115}, 102390 (2020).

\bibitem{agostinelli_geant4simulation_2003}
S.~Agostinelli et~al., ({GEANT4}),
  \href{https://doi.org/10.1016/S0168-9002(03)01368-8}{{{Geant4—a simulation
  toolkit}}}, Nucl. Instrum. Methods Phys. Res. Sect. A {\bf 506}, 250--303
  (2003).

\bibitem{BASUNIA201569}
M.~S. Basunia,
  \href{https://doi.org/https://doi.org/10.1016/j.nds.2015.07.002}{{{Nuclear
  data sheets for a = 22}}}, Nuclear Data Sheets {\bf 127}, 69--190 (2015).

\bibitem{CHEN20171}
J.~Chen,
  \href{https://doi.org/https://doi.org/10.1016/j.nds.2017.02.001}{{{Nuclear
  data sheets for a=40}}}, Nuclear Data Sheets {\bf 140}, 1--376 (2017).

\bibitem{NISTXray2005}
R.~Deslattes et~al.,
  \href{https://doi.org/https://dx.doi.org/10.18434/T4859Z}{{{X-ray transition
  energies}}}, (2005).
\newblock Accessed on 2023 Nov. 23.

\bibitem{gaiser_charmonium_1982}
J.~E. Gaiser,
  \href{https://www.slac.stanford.edu/cgi-bin/getdoc/slac-r-255.pdf}{{{Charmonium
  {Spectroscopy} from {Radiative} {Decays} of the {J}/{Psi} and
  {Psi}-{Prime}}}},
\newblock ph.{D}. {Thesis}, Stanford Linear Accelerator Center, Stanford
  University, (1982).
\newblock The function is defined on p.178.

\bibitem{choi_waveform_2024}
J.~J. Choi et~al., \href{https://doi.org/10.48550/arXiv.2402.17125}{{{Waveform
  {Simulation} for {Scintillation} {Characteristics} of {NaI}({Tl})
  {Crystal}}}}, \href{http://arxiv.org/abs/2402.17125}{{\tt arXiv:2402.17125}}
  (2024).

\bibitem{rasco_nonlinear_2015}
B.~C. Rasco et~al., \href{https://doi.org/10.1016/j.nima.2015.03.087}{{{The
  nonlinear light output of {NaI}({Tl}) detectors in the {Modular} {Total}
  {Absorption} {Spectrometer}}}}, Nucl. Instrum. Methods Phys. Res. Sect. A
  {\bf 788}, 137--145 (2015).

\bibitem{cano-ott_monte_1999}
D.~Cano-Ott et~al.,
  \href{https://doi.org/10.1016/S0168-9002(99)00217-X}{{{Monte {Carlo}
  simulation of the response of a large {NaI}({Tl})total absorption
  spectrometer for β-decay studies}}}, Nucl. Instrum. Methods Phys. Res. Sect.
  A {\bf 430}, 333--347 (1999).

\end{thebibliography}\endgroup
